\documentclass[conference]{IEEEtran}
\IEEEoverridecommandlockouts

\usepackage{cite}
\usepackage{amsmath,amssymb,amsfonts}
\usepackage{algorithmic}
\usepackage{graphicx}
\usepackage{textcomp}
\usepackage{xcolor}
\usepackage{multirow}
\usepackage{fancyhdr}
\def\BibTeX{{\rm B\kern-.05em{\sc i\kern-.025em b}\kern-.08em
    T\kern-.1667em\lower.7ex\hbox{E}\kern-.125emX}}

\begin{document}

\title{Digital Twin-Guided Energy Management over Real-Time Pub/Sub Protocol in 6G Smart Cities\\}

\author{\IEEEauthorblockN{Kubra Duran\IEEEauthorrefmark{1}, Lal Verda Cakir\IEEEauthorrefmark{1}\IEEEauthorrefmark{2}, Sana Ullah Jan\IEEEauthorrefmark{1}, Kerem Gursu\IEEEauthorrefmark{2} and Berk Canberk\IEEEauthorrefmark{1}}
 \ \
\IEEEauthorblockA{\IEEEauthorrefmark{1}School of Computing, Engineering and The Built Environment, Edinburgh Napier University, UK}
\IEEEauthorblockA{\IEEEauthorrefmark{2}BTS Group, Istanbul, Turkey} \
		Emails: \{kubra.duran, lal.cakir, s.jan, b.canberk\}@napier.ac.uk, kerem.gursu@btsgrp.com}

\maketitle
\fancypagestyle{firstpage}{%
  \fancyhf{}
  \fancyhead[L]{\normalsize Accepted by 2025 IEEE Global Communications Conference (GLOBECOM), \copyright~2025 IEEE}
  \renewcommand{\headrulewidth}{0pt} 
}
\thispagestyle{firstpage}

\begin{abstract} 
Although the emergence of 6G IoT networks has accelerated the deployment of enhanced smart city services, the resource limitations of IoT devices remain as a significant problem. Given this limitation, meeting the low-latency service requirement of 6G networks becomes even more challenging. However, existing 6G IoT management strategies lack real-time operation and mostly rely on discrete actions, which are insufficient to optimise energy consumption. To address these, in this study, we propose a Digital Twin (DT)-guided energy management framework to jointly handle the low latency and energy efficiency challenges in 6G IoT networks. In this framework, we provide the twin models through a distributed overlay network and handle the dynamic updates between the data layer and the upper layers of the DT over the Real-Time Publish Subscribe (RTPS) protocol. We also design a Reinforcement Learning (RL) engine with a novel formulated reward function to provide optimal data update times for each of the IoT devices. The RL engine receives a diverse set of environment states from the What-if engine and runs Deep Deterministic Policy Gradient (DDPG) to output continuous actions to the IoT devices. Based on our simulation results, we observe that the proposed framework achieves a 37\% improvement in 95th percentile latency and a 30\% reduction in energy consumption compared to the existing literature.  

\end{abstract}


\section{Introduction}

The evolution of 6G-enabled Internet of Things (IoT) networks is set to transform real-time applications in smart cities. This progress necessitates data-driven communication and intelligent decision-making capabilities  \cite{intro1, bc1, bc3}. As emerging technologies, Digital Twin (DT) and Artificial Intelligence (AI) play a crucial role to enhance communication efficiency by ensuring seamless and real-time applications. In addition, among the key enablers, Data Distribution Service (DDS) provide a communication infrastructure for time-sensitive applications. However, the proper and robust application of these enablers require system-oriented design to handle specific challenges. With this motivation, we focus on the following challenges for the full potential of 6G IoT networks for smart city services.

\begin{itemize}
    \item \textit{Low latency in smart city management:} Real-time responsiveness is crucial for 6G smart city management in which multiple smart city services are provided. At this point, it is essential to meet service demands without interfering with each other while avoiding additional increased latency in other services.
    \item \textit{Energy efficiency of IoT devices:} 6G requires massive connectivity, but the continuous operations needed to meet this demand pose a challenge for battery-constrained devices. At this point, the energy consumption at both the device and network levels requires further developments to serve energy-efficient operations.  
\end{itemize}

\begin{figure*}
\centerline{\includegraphics[width=.9\textwidth]{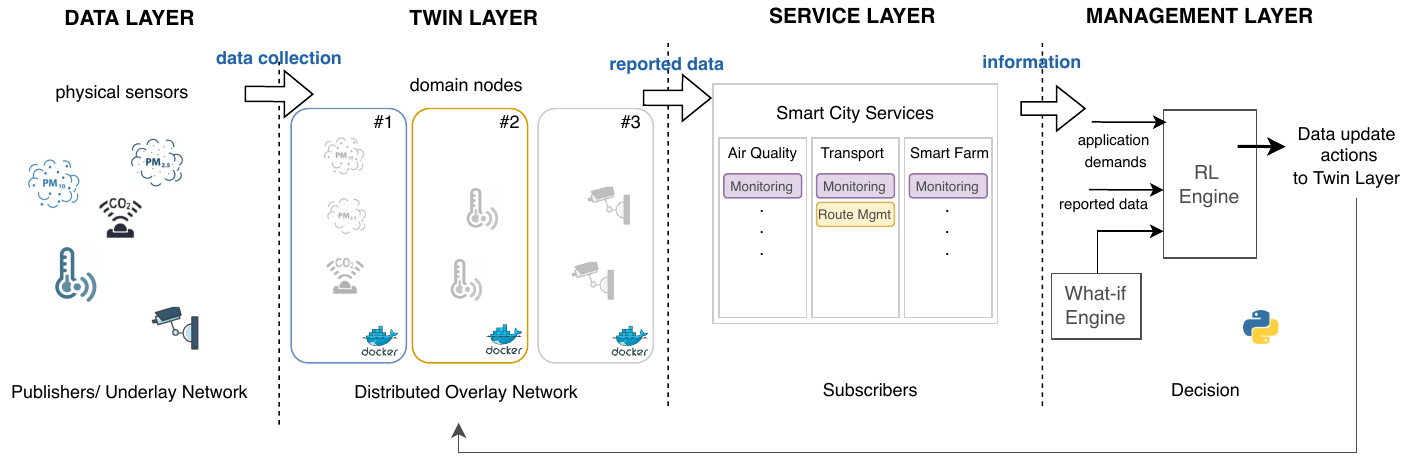}}
\caption{4-layered architecture of the proposed Digital Twin-guided energy management framework.}
\label{fig1}
\end{figure*}

\section{State of the Art}

The performance requirements of different smart city applications must be met for successive 6G integration. For this purpose, \cite{smart_city_network} proposes the utilisation of overlay networks in smart city services. In this way, the IoT devices are partitioned based on the domain they belong to. Moreover, \cite{iot_overlay_network} introduces a tiered overlay network approach to manage the distributed smart city services. In addition, the management protocol for battery-constrained devices should be lightweight and have minimal overhead for successive 6G smart city applications \cite{iot_management_protocol, bc2}. At this point, selecting the protocol parameters is important for the delay performance \cite{bc4}. For instance, in \cite{broker}, an MQTT scheme with different QoS levels performs the minimum delay with the provided acknowledgement. 
Although a retransmission mechanism is proposed in \cite{energy_concious_access_point} to prevent this, it leads to additional energy consumption during the transmission. In another study, \cite{ccnc}, a Hidden Markov Model (HMM)-based data update mechanism is proposed to minimise the latency in network discovery operation while considering the resource consumption. Similarly, \cite{tgcn} proposes an energy-efficient discovery while \cite{rewire} proposes software-defined adaptive protocol strategies by managing resource usage.

At this point, all these data update operations put strain on battery-constrained IoT devices even if they are provided over DTs \cite{gentwin}. To address this issue, in \cite{iot_scheduling}, the set of IoT devices to send data is decided by a Deep Q-Network (DQN), which operates on the discrete action space. However, in the context of DT, this approach may fall short of supporting the continuous operation of DTs. Because the timeliness criteria must be met to ensure the responsiveness of the DT \cite{timeliness}. 
Regarding this issue, \cite{ecco} also proposes a Deep Reinforcement Learning (DRL)-based energy control and computation offloading (ECCO) algorithm. It utilises Proximal Policy Optimisation (PPO) to optimise offloading decisions by deciding whether to process data locally or offload to an edge server. Similarly, \cite{d2lieo} uses Convolutional Neural Networks (CNNs) and proposes a Deep Dynamic Learning IoT-Edge Offloading (D2LIEO) algorithm to optimise the offloading decisions and thus optimise the energy consumption. 

As summarised above, several efforts have been made to tackle the low latency and energy efficiency challenges in 6G smart city networks. However, most of these mainly focus on offloading strategies to decrease energy consumption and communication latency.
Therefore, there is still a research gap for high-performance DT applications in future networks regarding latency and energy efficiency challenges. In this paper, our research question for this study is, \textit{“How can we effectively manage smart city services by (i) preserving the energy, and (ii) meeting the low latency requirements of future 6G IoT networks?”} To address this, we propose a DT-guided energy management framework that operates over the DDS. With this framework, we provide the interaction and data flow between the data layer and the applications over the Real-Time Publish Subscribe Protocol (RTPS). In the management layer, we utilise a Reinforcement Learning (RL)-based action recommendation with a novel reward function to predict optimal times for IoT devices to switch their active states and send the data to the relevant topics through the DDS. We use the Deep Deterministic Policy Gradient (DDPG) algorithm as it can generate continuous action values, which are suitable for the DT decision-making process. We list our contributions below:

\begin{itemize}

    \item  We create the twin models as a distributed overlay network by using the Data Distribution Service (DDS). We represent the twins with nodes, each assigned a domain ID based on the smart city service to which they are connected. All these nodes are interconnected by logical links (overlay links), and each logical link represents one or more physical links (underlay links) from the physical environment. 
    
    \item We provide Real-Time Publish Subscribe (RTPS)-based data distribution to handle dynamic updates between the data layer and the upper layers of the digital twin. For this, we implement RTPS discovery and data exchange operations set through the twin layer. 

    \item We design a Reinforcement Learning (RL) engine to predict optimal times for IoT devices to send up-to-date data to respective topics. The engine implements Deep Deterministic Policy Gradient (DDPG) with a novel reward function formulated for this study. As a result, the actions taken by the RL engine trigger the wake-up mechanism of the IoT sensors to initiate the data updates. 
    
\end{itemize}

The remainder of the article is organized as follows: Section III explains the proposed Digital Twin-guided energy management framework. Section IV details the performance evaluation. Section V presents the conclusion of the paper.

\section{Digital Twin-Guided Energy Management Framework}

The proposed digital twin-guided energy management framework consists of four distinct layers; the Data Layer, the Twin Layer, the Service Layer and the Management Layer.

\begin{table*}[]
\centering
\caption{RTPS Operations Set Through the Twin Layer}
\begin{tabular}{l|l|l}
\hline
Role                                   & \multicolumn{1}{c|}{Primitive} & \multicolumn{1}{c}{Description}                                                   \\ \hline\hline
\multirow{2}{*}{participant discovery} & \textit{data (available)}      & Indicates a participant is available.                                             \\
                                       & \textit{info}                  & Includes source information for discovery with the unique ID of all twin objects. \\ \cline{1-1}
\multirow{3}{*}{endpoint discovery}    & \textit{data (W)}              & Publisher announces the presence of a writer endpoint.                            \\
                                       & \textit{data (R)}              & Subscriber announces the presence of a reader endpoint.                           \\
                                       & \textit{heartbeat}             & Indicates a participant is active in the network.                                 \\ \cline{1-1}
\multirow{3}{*}{data exchange}         & \textit{publish}               & Used to disseminate events by a publisher.                                        \\
                                       & \textit{subscribe}             & Used to collect information for interested topics by subscribers.                 \\
                                       & \textit{state synch.}          & To align timestamps of participants for real-time coordination                   
\end{tabular}
\label{tab1}
\end{table*}

\subsection{Data Layer} 
This layer stands as the main data provider within the framework. All the physical publishers, also known as IoT devices for our study, are located in this layer. These devices deliver their readings, energy consumption metrics, battery levels, and device operational states (idle, active or sleep). With the utilization of a publisher/subscriber architecture, the dynamic addition or removal of devices is supported in this layer. Furthermore, this layer ensures that up-to-date information can be accessed by the digital twin system through this layer. The collected data from publishers is transmitted to the Twin Layer.

\subsection{Twin Layer}
\subsubsection{Distributed Overlay Network Creation}
We create the distributed overlay network by using the \textit{domain participant} feature of DDS. Each domain participant within the DDS forms an overlay node. Each of these overlay nodes represent the twin model of a smart city entity in our case.  For this reason, we define three domain IDs, such as \textit{service-based domain IDs}, such as \textit{AIR\_Q\_DOMAIN}, \textit{TRANSPORT\_DOMAIN}, \textit{SMART\_FARM\_DOMAIN}. We deploy each domain with an individual docker container over the twin layer. Furthermore, the overlay network is updated dynamically to reflect the changes in the Data Layer. This is done via topic-based synchronization in which each domain participant subscribes to relevant topics. And this ensures that updates propagate across the overlay network efficiently. 

\subsubsection{RTPS-based Data Distribution}
We use RTPS to maintain the data flow between the Data Layer, Twin Layer and the Application Layer. The data flow consists of updates for \textit{data exchange} (delivering and collecting the up-to-date data) as well as \textit{discovery} information.  As given in Table \ref{tab1}, we use two types of discovery, \textit{participant discovery}, and \textit{endpoint discovery}. With the former one, we allow the smart city services to integrate into the network via distributed overlay nodes. For this, the availability of a smart city service (participant) and the required \textit{info} is send to the other participants. With endpoint discovery, 
we allow the twin models to integrate into the network via distributed overlay nodes. For this, we allow data flow between \textit{writers} (twins) and \textit{readers} (applications) and also \textit{heartbeat} messages to highlight active network entities. With these discovery mechanisms, when a new smart city entity (let's say a new traffic sensor) is introduced, it is automatically registered to the related smart city domain which is transport. If an existing node goes offline, the overlay network self-updates to reroute messages accordingly. 

In addition, we maintain \textit{data exchange} operations for submitting up-to-date data readings (\textit{publish}) and the collection of data (\textit{subscribe}) when new data is present in the domains. \textit{State synchronization} is also performed for the alignment of participant timestamps for real-time coordination between the Data Layer and Twin Layer. We give all the operations we maintain for the data flow between the Data Layer and the upper layer of the digital twin during our implementations in Table \ref{tab1}. 

\begin{figure*}
\centerline{\includegraphics[width=.85\textwidth]{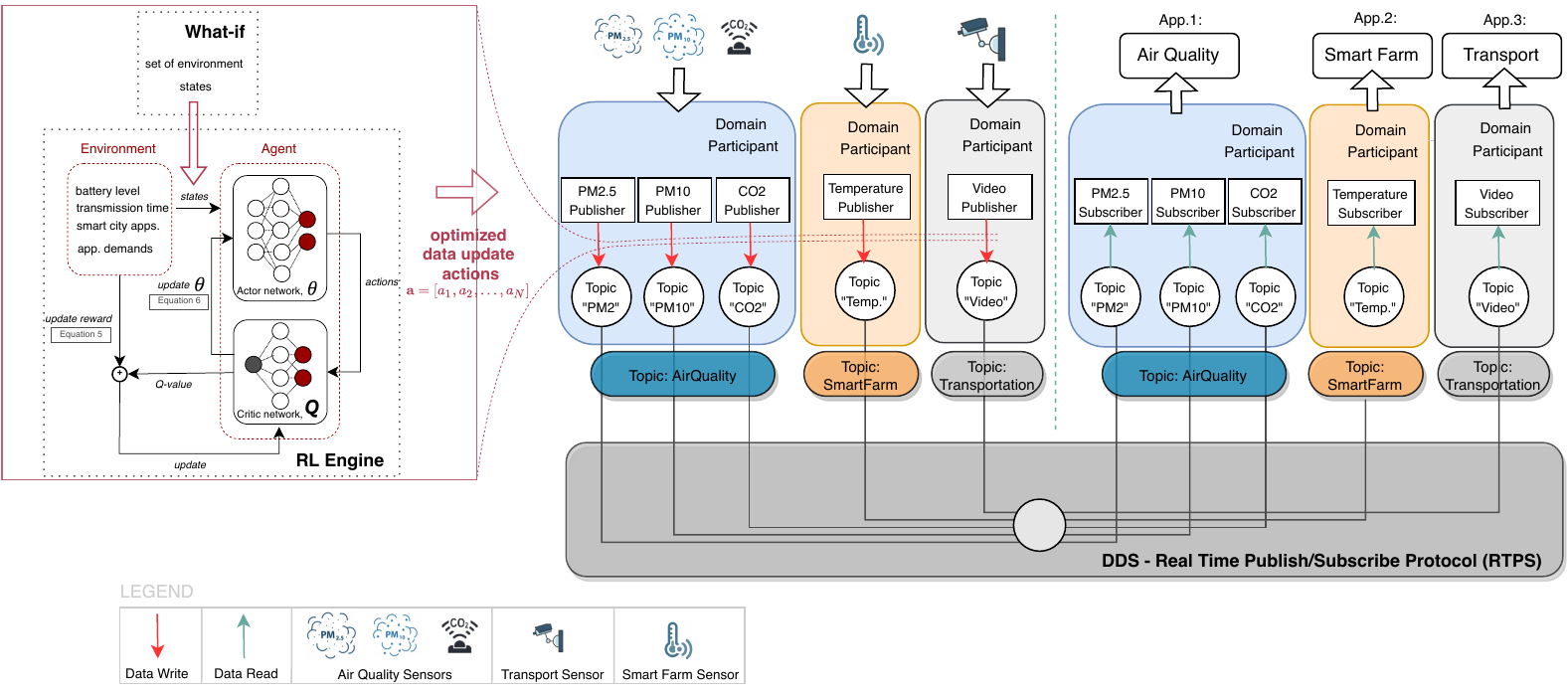}}
\caption{Management Layer, Twin Layer, and Service Layer interactions over RTPS.}
\label{fig2}
\end{figure*}

\subsection{Service Layer}
All smart city applications are managed through this layer. The monitoring and management of service-specific functions, such as route management for transport services or emission tracking for air quality services, are all executed within this layer. In our design, we focus on three distinct smart city services: Air Quality, Transportation, and Smart Farm by using the specific datasets given in Section IV.

\subsection{Management Layer} 
This layer consists of an RL engine, and a What-if engine. In the What-if engine, we provide a divergent set of environmental states to enhance the performance of RL engine. In the RL engine, we implement DDPG algorithm to learn optimal policies. The RL model consists of four tuples $\{S, A, R, S'\}$, where $S$ is the current state space, including the current battery level readings of the IoT devices, $A$ is the action space, $R$ is the reward set, and $S'$ is the predicted state space. We explain the details below: 

    \begin{itemize}
    \item \textit{State, S:} The state vector contains the battery levels of the devices and the last transmission times of each device to balance energy and timeliness constraints. We can indicate an example of state space like,  \(s = [\text{battery}_1, \text{battery}_2, \text{last\_transmission}_1, \dots]\)

    \item \textit{Action Space, A:} Each action $a_i$ represents the time (in minutes) within the 24-hour period when device $i$ sends its data. The continuous action vector \(\mathbf{a} = [a_1, a_2, \dots, a_N]\),  where each $a_i$ is the sending time for sensor i. Therefore, we can indicate the actions as \(a_i \in [0, 1440]\), where $a_i = 0$ means that the device will transmit at midnight, and $a_i = 1440$ means the device transmits just before midnight of the next day. To ensure no two transmissions from the same device are scheduled within a certain interval \(\Delta t_{\text{min}}\), we define the constraint while creating the next actions for the same devices as given below:
    
    \begin{equation}
    a_i' = \max(a_i + \Delta t_{\text{min}}, \text{last transmission time})
    \end{equation}

    \item \textit{Reward:} As our main target in this study is to preserve energy and at the same time maintain minimum delays, we formulate a reward function by considering remaining battery levels and criteria for timeliness. For this, we consider obtaining a positive reward value for the actions that result in higher battery levels and also timely delivery. We calculate the energy reward by using,
\begin{equation}
        R_{\text{energy}} = \sum_{i=1}^{N} B_i(a_i)
        \label{eq3}
\end{equation}

where, $B_i(a_i)$ indicates the resulting battery level at the time when $a_i$ is applied. Furthermore, we calculate the timeliness reward  by using, 

\begin{equation}
        R_{\text{timeliness}} = -\sum_{i=1}^{N} \max\big(0, D_i - D_\text{threshold})
        \label{eq4}
\end{equation}

In (\ref{eq4}), we check if the communication delay for the respective sensor $i$ falls below the delay threshold ($D_\text{threshold}$) or not. In this case, we can observe two scenarios, such as:

        \begin{itemize}
            \item If $D_i > D_{threshold}$, the reward value will be equal to the second term that is $D_i - D_{\text{threshold}}$. As it will corrupt the timeliness criteria, an additional penalty term will appear in the reward function as the second term. 
            \item If $D_i \leq D_{threshold}$, the reward value will be equal to the first term, meaning that the action applied meets with the timeliness criteria.
        \end{itemize}

Furthermore, we define a penalty for consecutive transmissions that don't meet with the action constraint as,

\begin{equation}
        R_{\text{consecutive}} = -\lambda \cdot \sum_{i=1}^{N} \max\big(0, \Delta t_{\text{min}} - (a_i' - a_i)\big)
\end{equation}

, where $\lambda$ stands for the penalty factor for violating the consecutive transmission constraint. Here, we use this scaling factor to indicate the relation of consecutive transmissions with the battery usages. More specifically, each consecutive transmission that occurs by violating the minimum interval we set means higher battery usage in our scenario. As we set this minimum interval dynamically during the RL algorithm run, each transmission violating this interval should have a negative impact for the reward function. That's why we employ such a penalty factor. Furthermore, we form the total reward function by using the energy reward, timeliness reward and the consecutive transmission penalty values. We use the total reward as,

\begin{equation}
        R_{\text{Total}} = R_{\text{energy}} +   R_{\text{timeliness}} +  R_{\text{consecutive}}
        \label{eq5}
\end{equation}

\item \textit{Actor Network ($\theta$):} It is the policy network ($\pi$) that learns the policy to select optimal actions. For this, it takes the current state of the IoT network and outputs a continuous action by representing the transmission times of IoT devices. In the run of the DDPG algorithm, it updates the Q values based on the environment and learning rate values to output the action-value function under the optimal policy. For example, in our scenario, it evaluates if the applied message update period results in longer battery lifetimes.

\item \textit{Critic network ($Q$):} It is the value function network that evaluates the actions taken in a given state. In our scenario, the critic network evaluates if the decided data transmission times for particular sensors are optimal for energy consumption and timeliness constraints. For this, it estimates the Q-value as the long-term reward evaluation.

\item \textit{Actor Target Network ($\theta_{target}$):} It is used to generate the target action $a_{target}^{'}$ for the next state $s'$. In our scenario, this network produces the candidate values for the data transmission times which result in preserving more battery levels in percentage.

\item \textit{Critic Target Network ($Q_{target}$):} It is used to compute the target Q-value, $Q_{target}(s',a_{target}^{'})$, which is required to train the critic network. Therefore, we can say that the critic target network is important for minimization of energy consumption and thus the efficient management of battery levels in our scenario.

\item \textit{Replay Buffer:} This is responsible for storing experiences in the form of transitions.

\item \textit{Loss Function:} The main aim of this function is to focus on improving the policy to achieve higher rewards in future states. 
We calculate the loss as a mean squared error (MSE) loss as given below:

        \begin{equation} 
      Loss = \frac{1}{b} \sum (Q_{target}(s',a_{target}^{'}) - Q(s, a))^2
        \end{equation}
, where $b$ is the batch-size for a single training iteration sampled from the replay buffer. 

In the running of the DDPG, states are observed from the Twin Layer by using the virtual metadata of IoT topology. Following this, the learner model takes action, $a_i$ with the 
 $\epsilon$-greedy mechanism by randomly choosing the exploration and exploitation phases. Thus, the agent takes a random action at a given time with the probability of $\epsilon$ or (1 $-$$\epsilon$). Next, the time slot is moved to $t+1$. 
 \end{itemize}

\section{Performance Evaluation}
In this section, we investigate the performance of the proposed Digital Twin-guided energy management framework. For this, we measure (i) the 95th percentile latency against an increasing volume of communicated data, and (ii) the total energy consumption of the smart city network with the increasing number of publishers. For our experiments, we utilise 1-service and 3-service smart city scenarios. In this context, the 1-service scenario indicates that the Service Layer of the DT implements only one of the smart city services, whereas the 3-service scenario refers to the implementation of all three smart city services (air quality, transport, and smart farming) within the Service Layer. We create our simulation environment in MatlabR2024b\textsuperscript{\copyright} by using the data from \cite{air}, \cite{farm}, and \cite{transport}. We use Python for DDPG implementation and to execute our testing scenarios. We also use WireShark\textsuperscript{\copyright} with the version of 4.4.3 for live capturing of the data flow through RTPS. All the simulation parameters are given in Table \ref{tab:sim}.

\begin{table}[thpb!]
    \centering
\caption{Simulation Parameters} 
\centering 
    \begin{tabular}{l c} 
\hline 
Parameters & Values \\ [0.5ex]
\hline\hline 
Total number of publishers & [0, 600]\\
Learning rate & [0.001, 0.2]  \\
Optimizer function & sgdm  \\
Discount factor  & 0.8 \\
$D_{\text{threshold}}$ & 180ms \\
Batch size & \ 256\\
Penalty factor, $\lambda$ & \ 0.3\\
Confidence interval  & 95\% \\
\hline 
    \end{tabular}
\label{tab:sim}
\end{table}

\begin{figure}
\centerline{\includegraphics[width=.42\textwidth]{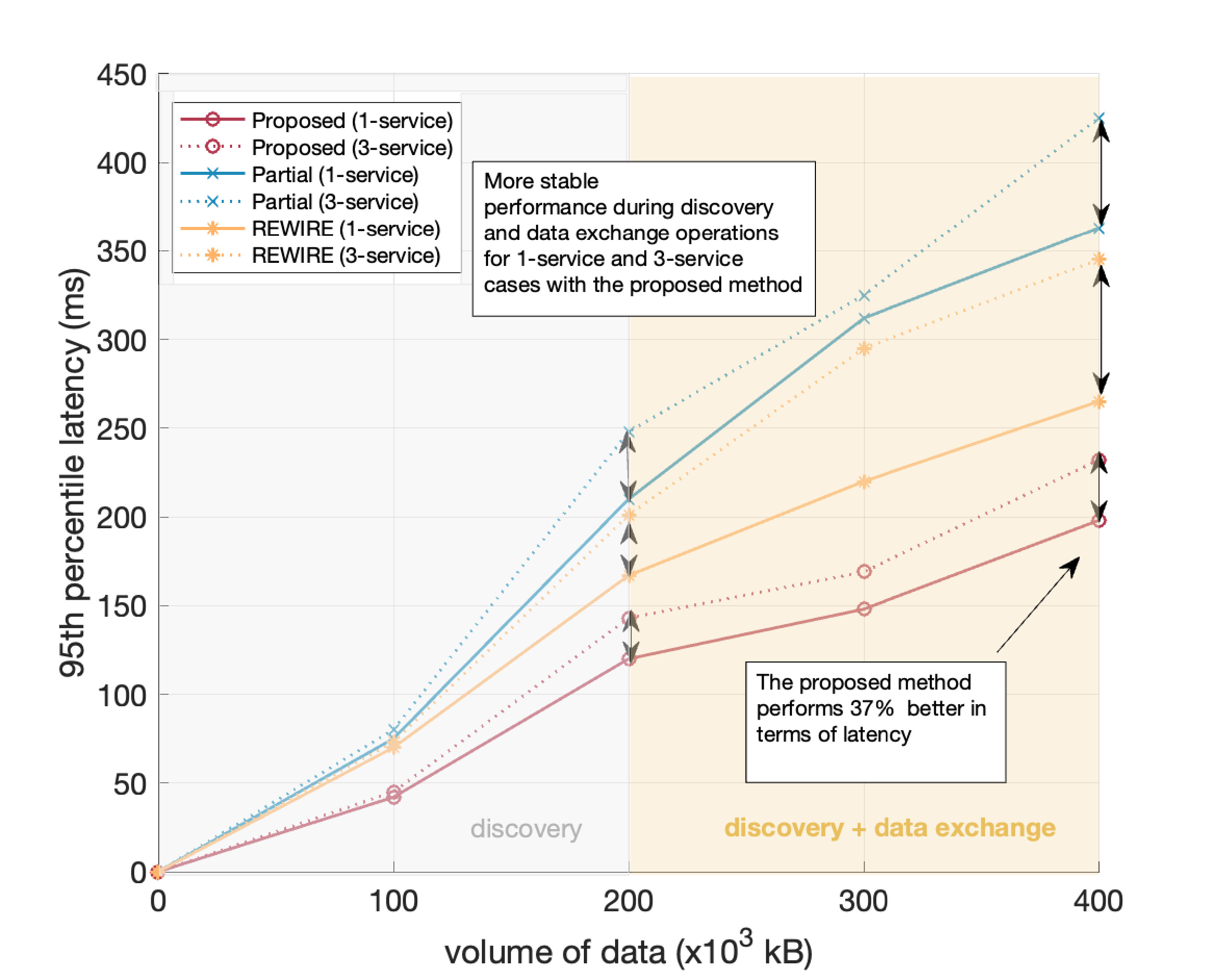}}
\caption{95th percentile delay vs increasing volume of data communicated for 1-service and 3-service scenarios.}
\label{latency}
\end{figure}

In the first set of our experiments, we search for the effect of increasing volume of data communicated between the DT layers individually for 1- and 3-service smart city scenarios onto the communication latency. For this, we start with the smallest data amount in kBytes and increase it up to 400MB. In the first part of this information (up to 200MB), we allow only the network discovery by applying the discovery operation set of RTPS. In the second part (starting form 200MB, upon to 400MB), we also allow the data exchange operations between the DT layers. Moreover, in the information flow, we assume that the data amount is requested for each of the service applied within the Service Layer of Digital Twin. More specifically, in a 3-service scenario, each of the individual smart city services requests at most 400MB of information from the Twin Layer. We compare the performance of the proposed method with Partial\cite{ccnc} and REWIRE \cite{rewire} methods from the current literature. As given in Fig. \ref{latency}, we note that the proposed method performs more stable than the other methods for 1-service and 3-service smart city scenarios. Namely, during the discovery and data exchange operations, 95th percentile latency does not degrade too much with the proposed method. We show the degradations with the two-sided errors in the figure. Furthermore, we note that when the maximum volume of data communicated between the DT layers, the proposed method performs  37\% better in terms of 95th percentile latency. The main reason for this stems from the advantage of overlay networking through the Twin Layer and serving the communication between the DT layers with the RTPS.

\begin{figure}
\centerline{\includegraphics[width=.4\textwidth]{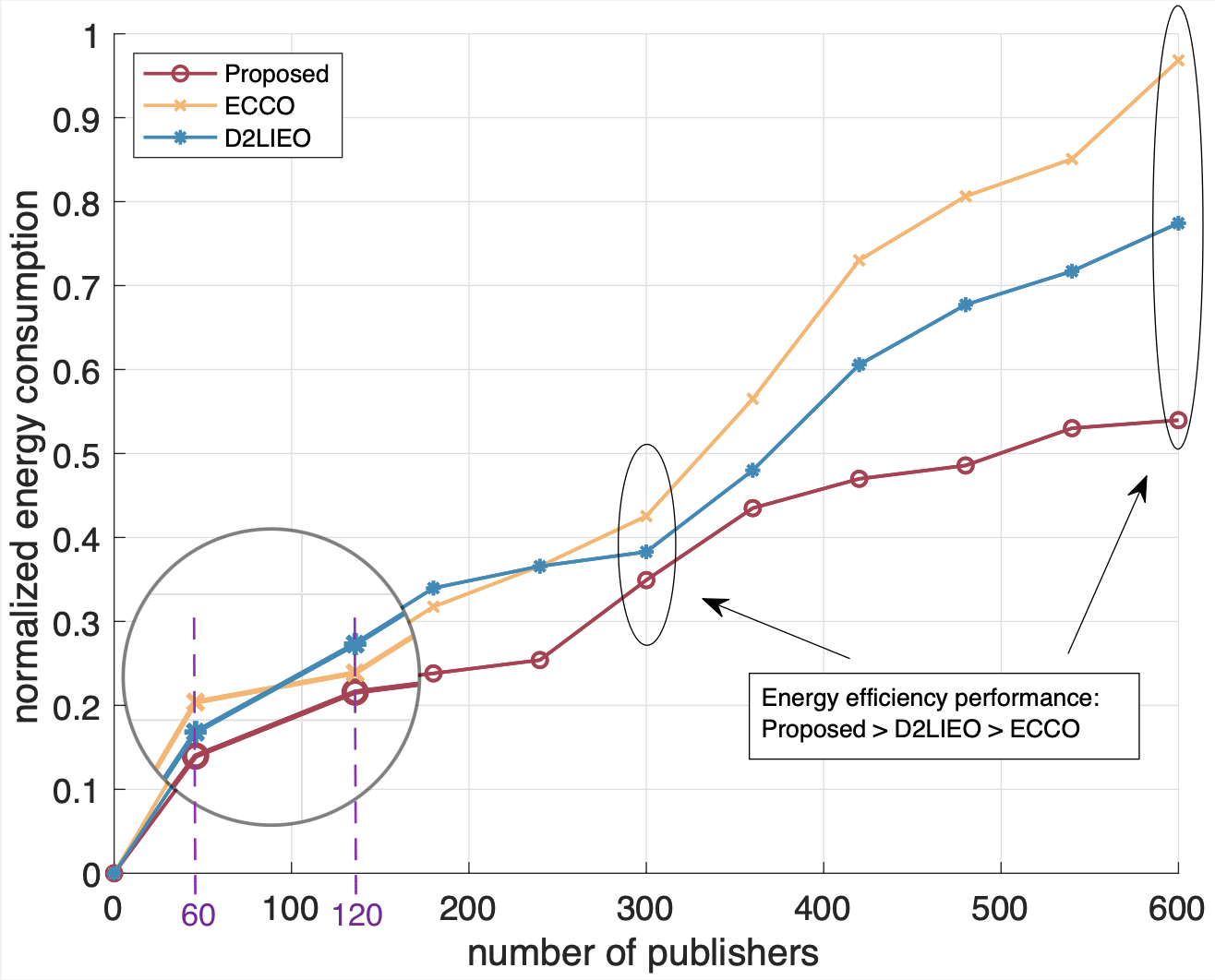}}
\caption{Energy consumption vs number of publishers with 3-service smart city scenario.}
\label{energy}
\end{figure}

In the second set of our experiments, we search for the energy efficiency of the proposed method by observing the total energy consumption values for the 3-service smart city scenario. For this, we evaluate the remaining battery levels of the devices in a cumulative manner, and assume that each decrease in the battery level is considered as the energy consumption value. We run the simulation up to 600 publishers in the smart city network. We compare the performance of the proposed method with ECCO \cite{ecco} and D2LIEO \cite{d2lieo} from the current literature. To clearly compare the energy efficiency of all the methods, we normalise the energy consumption value by referencing the maximum value produced by the ECCO. We give our observation results in Fig. \ref{energy}. As seen from the figure, when the number of publishers increase, the performances of the methods change two times (zoomed circle in the figure); but the most successive method (the proposed) does not change. Specifically, when the number of publishers reaches one hundred and twenty, we observe that the performance of ECCO and D2LIEO changes. Furthermore, they also change when the number of publishers exceeds two hundred and forty. This shows the stability of our proposed method, especially thanks to the outputs of the DDPG. For further observations, with the largest topology we run with 600 active publishers, we note that the proposed method surpasses the other ones with its energy efficiency performance by 30\%. D2LIEO follows the proposed method and ECCO results in the worst performance in terms of energy consumption. The main reason for this success is that the proposed method determines the optimum data update times for the IoT devices by eliminating the unnecessary active times for the devices thanks to the continuos action set with the DDPG.

\section{Conclusion}
In this study, we propose a Digital Twin (DT)-guided energy management framework to handle energy efficiency and low-latency requirements of 6G IoT networks. In our design, we provide (i) a distributed overlay network for the twin models, (ii) a real-time data flow between the DT layers over Real-Time Publish Subscribe (RTPS) Protocol, and (iii) a Reinforcement Learning (RL)-based, specifically with Deep Deterministic Policy Gradient (DDPG), what-if engine to output continuous actions for IoT devices. For RL, we design a novel reward function by considering energy efficiency and real-time constraints. According to the simulation results, our proposed method achieves a 30\% energy preservation and provides a 37\% improvement in latency.

\section*{Acknowledgment}
This work was partially supported by The Scientific and Technological Research Council of Turkey (TUBITAK) 1515 Frontier R\&D Laboratories Support Program for BTS Advanced AI Hub: BTS Autonomous Networks and Data Innovation Lab. Project 5239903.

\bibliographystyle{IEEEtran}
\bibliography{IEEEabrv, main}

\vspace{12pt}

\end{document}